\documentstyle[preprint,aps]{revtex}
\begin{document}
\draft
\begin{title} 
  {Excitations and phase segregation in a two component    
  Bose-Einstein condensate} 
\end{title}  
\author{A.S. Alexandrov$^{1}$  
  and V.V. Kabanov $^{1,2}$}  
\begin{address}  
  {$^{1}$ Department of Physics, Loughborough University, Loughborough LE11  
  3TU, United Kingdom,  
  $^{2}$Josef Stefan Institute 1001, Ljubljana, Slovenia} 
\end{address} 
\maketitle 
\begin{abstract} 
  Bogoliubov-de Gennes (BdG) equations and the excitation spectrum 
  of a two-component Bose-Einstein condensate (BEC) are derived with an 
  arbitrary interaction between bosons, including long-range and short 
  range forces. The nonconverting BEC mixture segregates into two phases 
  for some two-body interactions. Gross-Pitaevskii (GP) equations are 
  solved for the phase segregated BEC. A possibility of  
  boundary-surface and other localised excitations  is studied.   
\end{abstract} 
\narrowtext 
   
\noindent 
  Neutral and charged  (Coulomb) Bose-gases  became 
   recently  of particular  
  interest motivated by the observations of BEC in an alkali vapor  
  \cite{and,dav,bra,mew,hag}, and by the bipolaron theory of 
  high temperature superconductors \cite{alemot}, respectively. Their 
  theoretical understanding is based on    
   the Bogoliubov \cite{bog}  
  displacement transformation, separating a large 
  matrix element of the condensate field operator,  $\phi({\bf r},t)$,    
  from the total $\psi({\bf r},t)$ and treating the rest  
    $\tilde{\psi}({\bf r},t)=\psi({\bf r},t)-\phi({\bf r},t)$ as a  
    small fluctuation. The resulting (Gross-Pitaevskii) equation for the 
    condensate wave function $\phi({\bf r},t)$ provides the mean-field 
    description of the ground state and of the excitation 
    spectrum \cite{edw,bus,gol}. Beyond the mean-field approach the
  Bogoliubov-de 
    Gennes (BdG)-type  equations were derived \cite{alekabbee}, 
    describing eigenstates of the 'supra'condensate bosons.   
   
  In this letter we  extend the Bogoliubov theory to the two-component 
   nonconverting condensate by deriving GP and BdG 
   equations and excitation spectrum with an arbitrary two-body interaction,
  and solving GP 
   equations for a phase-segregated BEC. Our motivation originates in the 
   recent experimental \cite{hag} and theoretical \cite{bus,gol}  studies of
  BEC of 
   $^{87}Rb$ atoms in two different hyperfine states, and also in an 
   observation \cite{alekab} that the condensation temperature of a
  $two$-component 
   charged Bose gas quantitatively   
   describes the superconducting critical temperature of many cuprates. 
  Differently from Ref. \cite{bus,gol} we 
   consider nonconverting components (such as different elements), and an  arbitrary 
   (rather than short-ranged) two-body interaction.  
   
  The  Hamiltonian  of the two-component ($1$ and $2$) mixture of bosons
  in an external  
  field  with the vector,  $A({\bf r},t)$, and scalar, $U_j({\bf r},t)$, 
  potentials   is given by 
   
  \begin{eqnarray} 
  H&=&\sum_{j=1,2}\int d{\bf r} \psi_j^\dagger({\bf r})\left[-{({\bf 
        \nabla}- 
  i q_j  
  {\bf A({\bf r},t)})^2 
  \over{2m_j}}+U_j({\bf r},t) 
  -\mu_j\right]\psi_j({\bf r})\cr 
  &+& 
  \frac 1 2 \sum_{j,j'}\int d{\bf r} \int d{\bf r'} V_{jj'}({\bf r} -  
  {\bf r'})\psi_j^\dagger({\bf r}) 
  \psi_j({\bf r})\psi_{j'}^\dagger({\bf r'})\psi_{j'}({\bf r'}), 
  \end{eqnarray} 
  where $m_j$ and $q_j$ are the mass, and the effective charge (if any) of the 
  boson $j$ ($\hbar=1$). 
  If the two-body interactions $V_{jj'}({\bf r})$ are weak, the occupation 
  numbers of one-particle states are not very much different from those in  
  the ideal Bose-gas. In particular the lowest energy state   
  remains to be macroscopically occupied and the corresponding   
  component of the field operator $\psi({\bf r})$ has anomalously large  
  matrix element between the ground states of the system containing  
  $N+1$ and $N$ bosons. Hence, it is convenient to consider a grand canonical  
  ansamble, introducing the chemical  
  potentials $\mu_j$ to deal with the anomalous averages $\phi({\bf r})$ 
  rather than with the off-diagonal matrix elements. Using the Bogoliubov 
  displacement transformation in the equation of  
  motion for the field operators 
  $\psi({\bf r},t)=\phi({\bf r},t)+\tilde{\psi}({\bf r},t)$ 
  and collecting $c-number$ terms of  
  $\phi$, and terms linear in the  $supracondensate$ boson operators 
  $\tilde \psi$ one obtains a set of the GP and  BdG-type equations 
  \cite{alekabbee,bus,gol}. The macroscopic condensate wave functions 
  (i.e the order parameters) obey two coupled GP equations  
  \begin{equation} 
  i {\partial \over {\partial t}}\phi_{j}({\bf r},t) =  
  \hat{h}_{j}\phi_{j} 
  ({\bf r},t)+ 
  \sum_{j'}\int d{\bf r'} V_{jj'}({\bf r} - {\bf r'})|\phi_{j'}({\bf r'},t)|^2 
  \phi_j({\bf r},t) 
  \end{equation} 
  with the single-particle Hamiltonian $\hat{h}_j=-({\bf\nabla}- 
  i q_j {\bf A}({\bf r},t))^2/2m_j+U_j({\bf r},t)-\mu_j$. The 
  supracondensate  
  wave-functions satisfy  four  BdG equations  
  \begin{eqnarray} 
  \sum_{j'}\int d{\bf r'}V_{jj'}({\bf r}-{\bf r'})[|\phi_{j'}({\bf r'},t)|^{2}
  u_{j}({\bf r},t)+\phi^{*}_{j}({\bf  
  r'},t)\phi_{j'}({\bf r},t)u_{j'}({\bf r'},t)+\nonumber \\
  \phi_{j}({\bf r'},t)\phi_{j'}({\bf r},t)v_{j'}({\bf r'},t)] 
  =i {\partial \over {\partial t}}u_{j}({\bf r},t)-  
  \hat{h}_{j}u_{j}({\bf r},t), 
  \end{eqnarray} 
  and  
  \begin{eqnarray} 
  \sum_{j'}\int d{\bf r'}V_{jj'}({\bf r}-{\bf r'})[|\phi_{j'}({\bf r'},t)|^{2}
  v_{j}({\bf r},t)+\phi_{j}({\bf  
  r'},t)\phi^{*}_{j'}({\bf r},t)v_{j'}({\bf r'},t)+\nonumber \\
  \phi^{*}_{j}({\bf r'},t)\phi^{*}_{j'}({\bf r},t)u_{j'}({\bf r'},t)] 
  =-i {\partial \over {\partial t}}v_{j}({\bf r},t)-  
  \hat{h}^{*}_{j}v_{j}({\bf r},t). 
  \end{eqnarray}  
  Here we have applied the linear Bogoliubov transformation  
  for $\tilde{\psi}$ 
  \begin{equation} 
  \tilde \psi_{j} ({\bf r},t)= \sum_{n}u_{nj}({\bf r},t)(\alpha_{n}+\beta_{n}) 
  +v_{nj}^{*}({\bf r},t)(\alpha_{n}^{\dagger}+\beta_{n}^{\dagger}), 
  \end{equation} 
  where $\alpha_{n}, \beta_{n}$  are bosonic   
  operators annihilating quasiparticles in the quantum state $n$.  
  There is a sum rule, 
  \begin{equation} 
  \sum_{n}[u_{nj}({\bf r},t)u^{*}_{nj}({\bf r'},t)-v_{nj}({\bf r},t)
  v^{*}_{nj}({\bf r'},t)]=\delta({\bf r}-{\bf r'}), 
  \end{equation} 
  which retains the Bose commutation relations for all operators.  
   
  In the homogeneous system with no external fields the excitation wave 
  functions are plane waves 
  \begin{equation} 
  u_{{\bf k},j}({\bf r},t)=u_{{\bf k},j}\exp[i{\bf k}\cdot {\bf r}-i E({\bf  
  k})t], 
  \end{equation} 
  \begin{equation}  
  v_{{\bf k},j}({\bf r},t)=v_{{\bf k},j}\exp[i{\bf k}\cdot {\bf r}-iE({\bf  
  k})t], 
  \end{equation} 
  The condensate wave function is $({\bf r},t)$ independent in this case, 
  $\phi_j({\bf r},t)\equiv\phi_j$. Solving two GP equations one obtains 
  the chemical potentials as 
  \begin{eqnarray} 
  \mu_1=Vn_1+Wn_2 \cr 
  \mu_2=Un_2+Wn_1, 
  \end{eqnarray} 
  and solving four BdG equations one determines the excitation spectrum, 
  $E({\bf k})$ from 
  \begin{equation} 
  Det\left  
  [\matrix{\xi_1({\bf k})-E({\bf k}) & V_{\bf k}\phi_1^2 & W_{\bf k}\phi_1
  \phi_2^{*} & W_{\bf k}\phi_1\phi_2\cr 
  V_{\bf k}\phi_1^{*2} & \xi_1({\bf k})+E({\bf k}) & W_{\bf k}\phi_1^{*} 
  \phi_2^{*} & W_{\bf k}\phi_1^{*}\phi_2\cr 
  W_{\bf k}\phi_1^{*}\phi_2 & W_{\bf k}\phi_1\phi_2 & \xi_2({\bf 
      k})-E({\bf k}) &  U_{\bf k}\phi_2^2 \cr 
   W_{\bf k}\phi_1^{*} 
  \phi_2^{*} & W_{\bf k}\phi_1\phi_2^{*} &   U_{\bf k}\phi_2^{*2}
  &\xi_2({\bf 
      k})+E({\bf k})\cr}\right]=0.   
  \end{equation} 
  Here $\xi_1({\bf k})=k^2/2m_1 + V_{\bf k}n_1$,$\xi_2({\bf k})=k^2/2m_2 
  + U_{\bf k}n_2$, and  
  $ V_{\bf k}, U_{\bf k}, W_{\bf k}$ are the Fourier components of 
  $V_{11}({\bf r}), V_{22}({\bf r})$ and $V_{12}({\bf r})$, 
  respectively, $V\equiv V_0$, $U \equiv U_0$, $W \equiv W_0$, 
  and  $n_{j}=|\phi_j|^2$ the condensate densities. There are 
  two brunches of excitations with the dispersion  
  \begin{equation} 
  E_{1,2}({\bf k})=2^{-1/2}\left(\epsilon_1^2({\bf k})+\epsilon_2^2({\bf 
      k}) \pm \sqrt{[\epsilon_1^2({\bf k})-\epsilon_2^2({\bf 
      k})]^2 + {4k^4\over{m_1m_2}} W_{\bf k}^2 n_1n_2} \right)^{1/2}, 
  \end{equation}  
  where $\epsilon_1({\bf k})=\sqrt{k^4/(4m_1^2)+k^2V_{\bf k}n_1/m_1}$ 
  and $\epsilon_2({\bf k})=\sqrt{k^4/(4m_2^2)+k^2U_{\bf k}n_2/m_2}$ 
  are Bogoliubov's modes of two components. 
  If the interactions are  short-ranged (hard-core) repulsions, so that  
  $V_{\bf k}=V$, $U_{\bf k}=U$, $W_{\bf k}=W$, the spectrum, Eq.(11) is that 
  of Ref. \cite{gol}. In the long-wave limit both brunches are sound-like 
  with the sound velocities 
  \begin{equation} 
  s_{1,2}=2^{-1/2}\left(Vn_1/m_1+Un_2/m_2 \pm \sqrt{[Vn_1/m_1 
  -Un_2/m_2]^2 + 4W^2 n_1n_2/(m_1m_2)} \right)^{1/2}. 
  \end{equation} 
  The lowest brunch becomes unstable ($s_2 < 0$) if $W> 
  \sqrt{UV}$. When $W=\sqrt{UV}$ this brunch is quadratic in the 
  long-wave limit 
  \begin{equation} 
  E_2({\bf k})={k^2\over{2(m_1m_2)^{1/2}}} 
  \sqrt{{Vn_1m_1+Un_2m_2\over{Vn_1m_2+Un_2m_1}}}, 
  \end{equation} 
  If $m_1=m_2=m$ it becomes 'collisionless' \cite{gol}, i.e 
  $E_2({\bf k})=k^2/2m$.  
   
  The finite-ranged interactions drastically change the whole 
  spectrum. In the extreme case of the long-range Coulomb interaction, 
  $V({\bf k})= 4\pi q_1^2/k^2$, $U({\bf k})=4\pi q_2^2/k^2$, and 
  $W({\bf k})=4 \pi q_1q_2/k^2$ the upper brunch is the geometric sum of 
  the familiar plasmon modes 
  \cite{fol} for $k \rightarrow 0$, 
  \begin{equation} 
  E_1({\bf k})=\sqrt{{4\pi q_1^2n_1\over{m_1}}+{4\pi q_2^2n_2\over{m_2}}}, 
  \end{equation} 
  while the lowest brunch is  
  \begin{equation} 
  E_2({\bf k})={k^2\over{2(m_1m_2)^{1/2}}} 
  \sqrt{{q_1^2n_1m_1+q_2^2n_2m_2\over{q_1^2n_1m_2+q_2^2n_2m_1}}}. 
  \end{equation} 
  Remarkably, this mode is 'collisionless' at $any$  charges of the components 
  if $m_1=m_2$, Fig.1.  Physically, it describes  neutral oscillations of 
  condensates, when a charge fluctuation in one component is 
  nullified by a charge fluctuation in another component, and the total 
  charge density does not fluctuate. We conclude that  while the 
  Coulomb Bose condensate is a superfluid (according to the Landau criterion),  
  mixture  of two Coulomb Bose condensates is not. In a more general 
  case the interaction might include both the  long-range repulsion and 
  the hard-core interaction as in the case of bipolarons or any other 
  preformed bosonic pairs \cite{alemot}. Combination of both 
  interactions, i.e.  $V_{\bf k}, U_{\bf k}, W_{\bf k} \propto 
  constant +1/k^2$ transforms the lowest quadratic mode into the 
  Bogoliubov sound. Hence, two component condensate of bipolarons is 
  a superfluid.  
   
  Finally, let us discuss the phase segregated state  of the 
  two-component nonconverting mixture with the hard-core interactions 
  when $W>\sqrt{UV}$\cite{remark}. In that case chemical potentials determined 
  in Eq.(9) are no longer correct. Minimizing the free energy with respect to 
  the equlibrium concentration we find the densities
  $n_1^{'}=n_1+\sqrt{U/V}n_2$ and 
  $n_2^{'}=n_2+\sqrt{V/U}n_1$ of two separated phases.
  The phase boundary is described by two coupled one-dimesional GP 
  equations, Eq.(2), 
  \begin{equation} 
  {d^2f_1\over{dx^2}}+f_1-f_1^3 -rf_1f_2^2=0, 
  \end{equation} 
  and 
  \begin{equation} 
  \kappa {d^2f_2\over{dx^2}}+f_2-f_2^3 -rf_2f_1^2=0. 
  \end{equation} 
  where we introduce two real dimensionless order parameters, 
  $f_1=\phi_1/\sqrt{n_1^{'}}$ and $f_2=\phi_2/\sqrt{n_2^{'}}$, and measure  
  length in units of the coherence length $\xi_1=(2m_1 Vn_1^{'})^{-1/2}$. 
  Parameter $r\equiv W/\sqrt{UV}$ is larger than 1, and 
  $\kappa =m_1Vn_1^{'}/m_2Un_2^{'}$ is the ratio of two 
  coherence lengths squared. One can solve these equations analytically 
  in the limit $\kappa \rightarrow 0$, where two coherence lengths 
  differ  significantly or in the case $r \rightarrow \infty$. 
  In the limit $\kappa \rightarrow 0$ the first term in the second 
  equation is negligiable, so that $f_2=0$ if $rf_1^2>1$, and 
  $f_2=\sqrt{1-rf_1^2}$ if $rf_1^2 <1$.  Substituting this solution 
  into the first equation and using $df_1/dx=F(f)$, one can reduce Eq.(16) 
  to an integerable first order differential equation. The solution 
  satisfying the boundary conditions, $f_1=1$ at $x=\infty$ and $f_1=0$ at 
  $x=-\infty$ is 
  \begin{equation} 
  f_1(x)=\Theta(x-x_1)\tanh(x/\sqrt{2})+
  \Theta(x_1-x){\sqrt{2}\over{(r+1)^{1/2}\cosh[(r-1)^{1/2}(x-x_2)]}}, 
  \end{equation} 
  \begin{equation} 
  f_2(x)=\Theta(x_1-x)\left(1-{2r\over
  {(r+1)\cosh^2[(r-1)^{1/2}(x-x_2)]}}\right)^{1/2}, 
  \end{equation} 
  where $\Theta(x)$ is the $\Theta$-function, $\tanh(x_1/\sqrt{2})=r^{-1/2}$
  and  
  $\cosh[(r-1)^{1/2}(x_1-x_2)]=\sqrt{2r/(r+1)}$.    
  Interestingly the $largest$ coherence length (in our case  
  $\xi_1$) determines profile of $both$ order parameters. The inter-component 
  mutual repulsion ($r$) only slightly changes the characteristic lengths, as 
  shown in Fig.2.  
   
  The quasiparticle eigenstates of the phase segregated condensate are 
  determined by the $inhomogeneous$ BdG equations, Eqs.(3,4). Here we 
  restrict the analysis by surface excitations  at the boundary 
  between two condensates in the limit $r \rightarrow \infty$. Substituting 
  Eqs.(18,19) into Eqs.(3,4) we obtain 4 coupled linear differential equations. 
  In the limit $r \rightarrow \infty$ the equations are decoupled in
  pairs, so that one can 
  solve only first two equations at $x>0$. Integration of the equations near zero
  provides the boundary condition $u_1(0)=v_1(0)=0$. Then a simple analysis shows
  that 
  there are no  excitations localised near the boundary  in that limit. 

  This conclusion seems to be quite general for any gapless Bose liquid. For
  example, one can 
   consider a single component hard-core Bose 
  gas with attractive impurity potential placed at the origin $x=0$, so that
  the GP equation is
  \begin{equation} 
  {d^2f\over{dx^2}}+f-f^3 +\alpha \delta(x)=0, 
  \end{equation} 
  where $\alpha>0$ is the (dimensionless) strength of the potential. The
  solution
  is $f(x)=\coth[(|x|+x_0)/\sqrt{2}]$ where 
  $\sinh (x_0/2^{3/2})=1/\alpha$. 
  Substituting this solution into the BdG equations Eqs.(3,4) one can see 
  that localised excitations do not appear even in that case. The condensate 
  density increases at $x=0$ and effectively screens out the attractive
  potential.
  However, localised excitations could appear in the gapped 
  Bose-liquid, like the charged Bose gas. In that case 
  they are formed below the plasma frequency.

  In conclusion, we have derived the
  Bogoliubov-de Gennes equations and the excitation spectrum 
  of the two-component Bose-Einstein condensate with an 
  arbitrary interaction between bosons, including long-range and short 
  range forces. Solving GP equations  for  segregated condensates we  found
  the boundary profile,
  and showed that localised 
  surface waves do not exist for strongly repulsive 
  components.  
   
  We greatly appreciate  enlightening discussions of the nonlinear GP
  equations with Alexander Veselov.
  This work has been supported by EPSRC UK, grant R46977.

\newpage 
   
  Figure captions. 
  Fig. 1. The excitation energy spectrum of the two-component BEC with 
  the long-range repulsive and hard-core interactions. In this plot 
  $n_1=n_2=n$, $m_1=m_2$ and $U=V=\omega_p$, $q_1=q_2=e$. Different curves
  correspond to the different
values of  $W/V=0.1$ (dashed line), $0.5$ (dotted line), and $0.95$
  (solid line),  respectively. Excitation energy is measured 
  in units of the  plasma frequency $\omega_p=4\pi n e^2/m$ and momentum $k$ is 
  measured  in invers screening length $q_s= (16\pi e^2 nm)^{1/4}$.
  
  Fig. 2. Density profiles of two phase-separated condensates near the 
  boundary for $r=2$ (solid line) and $r=10$ (dotted line), respectively. 
  

\begin{references}
\bibitem{remark}
  The condition $W>\sqrt{UV}$, where the lowest branch of the spectrum Eq.(12) 
  becomes unstable, coinsides with the condition, where compressibility 
  of the system becomes negative $|\partial \mu_i/\partial n_j| <0$. Phase 
  segregation also exists in the case of the $Coulomb + hard-core$ interaction.
   In that case the condition where the spectrum Eq.(11) becomes unstable 
  depends on the plasma frequency. It shows that the higher order 
  corrections to the chemical potential due to the Coulomb interaction  are 
  crucial
\end{references}
  
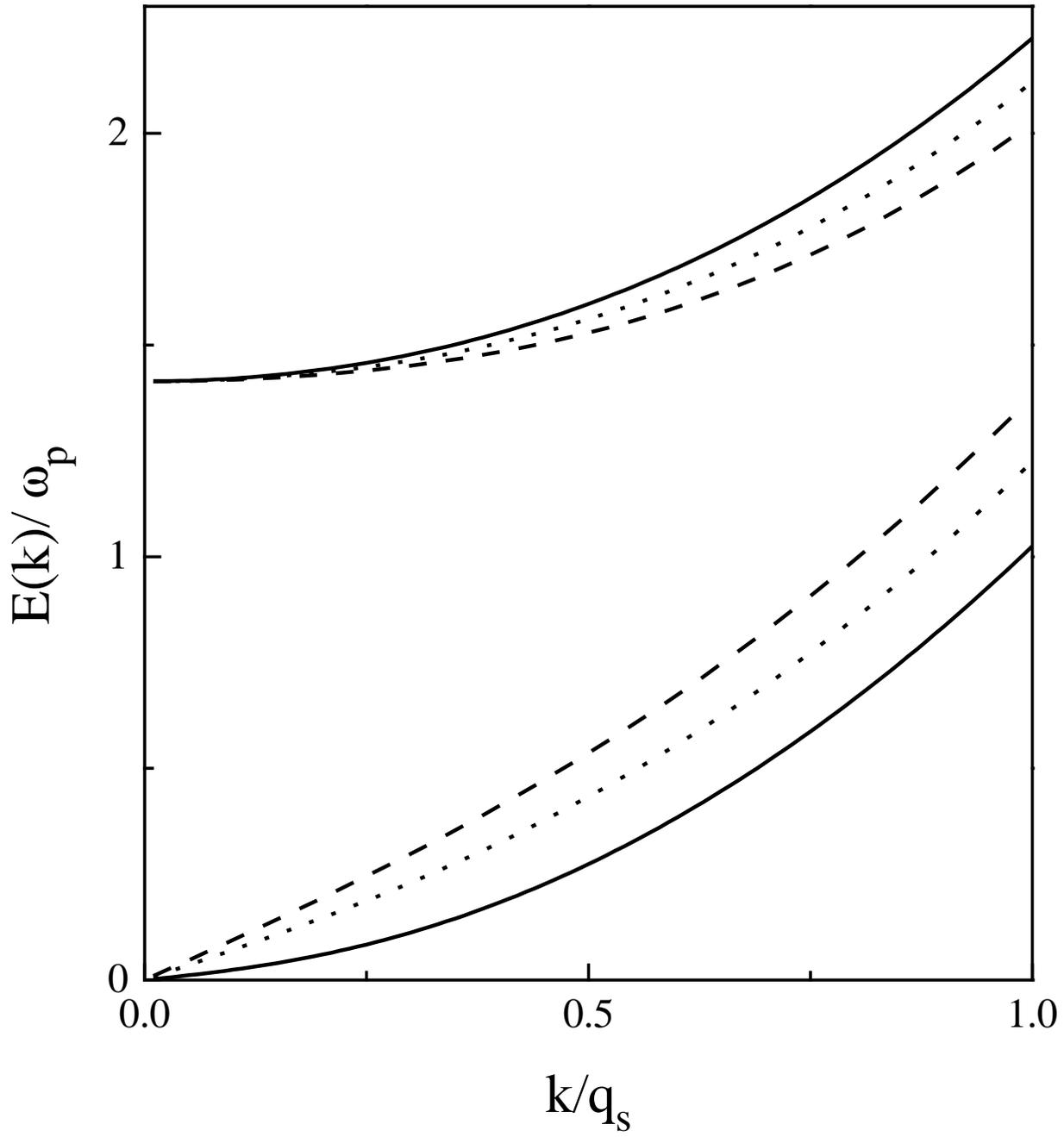
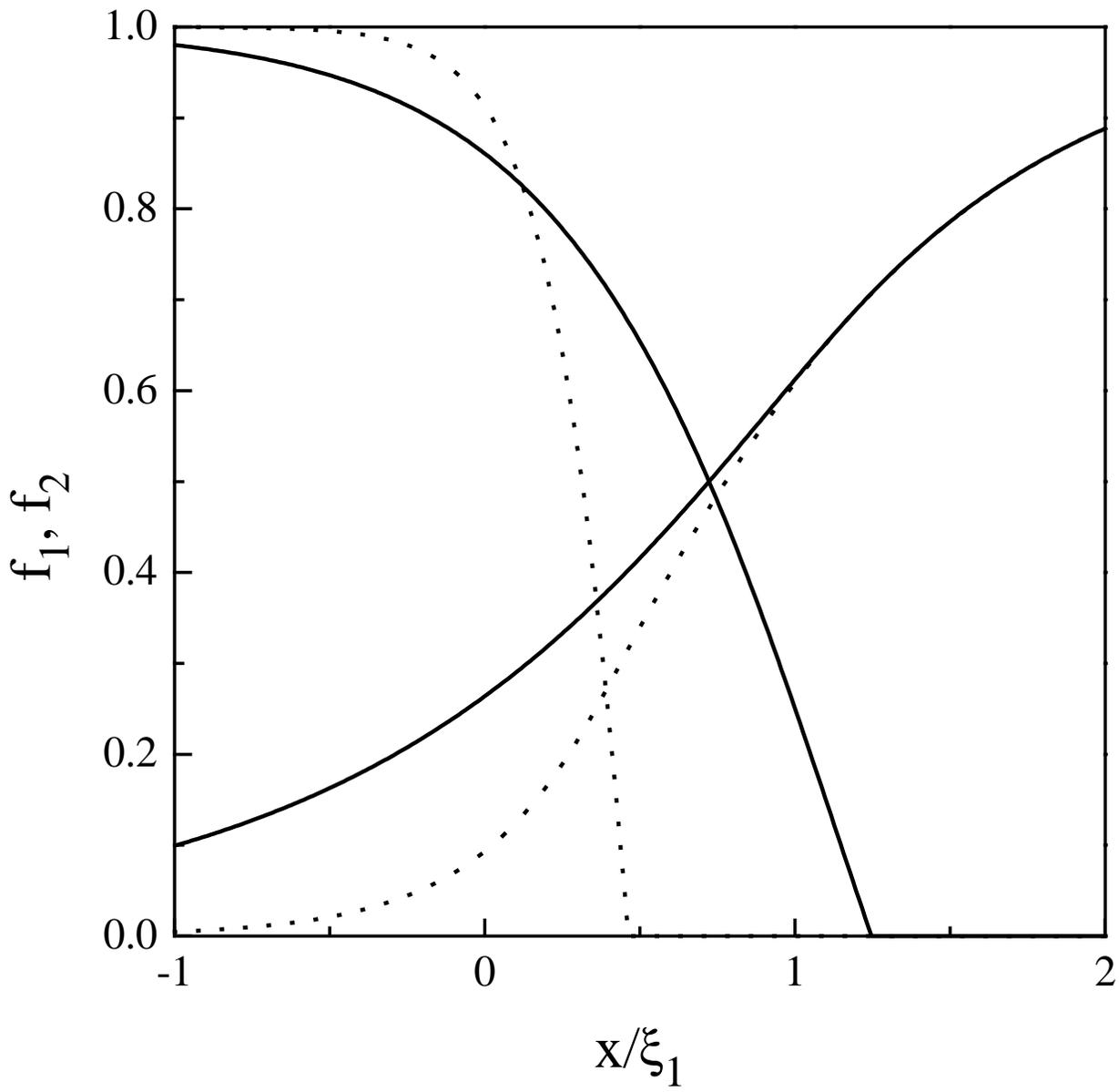
\end{document}